\documentclass[nofootinbib,a4paper,aps,twocolumn,prd,10pt,superscriptaddress]{revtex4}

\usepackage{graphicx}
\usepackage{amsmath}
\usepackage{amsfonts}
\usepackage{amsthm}
\usepackage{mathrsfs}
\usepackage{amssymb}
\usepackage{epsfig}
\usepackage{amscd}
\usepackage{dcolumn}
\usepackage{bm}
\usepackage{natbib}
\usepackage{url}
\usepackage{xspace}
\usepackage{kantlipsum}

\usepackage[normalem]{ulem}
\usepackage{appendix}
\usepackage{algorithm}
\usepackage{algorithmic}

\usepackage{siunitx}
\usepackage{paralist}
\usepackage{placeins}

\usepackage{xcolor}

\usepackage{comment}

\usepackage{hyperref}
\hypersetup{%
    ,urlcolor=blue
    ,citecolor=blue
    ,linkcolor=blue
    }

\def\be{\begin{equation}}
\def\ee{\end{equation}}
\def\bea{\begin{eqnarray}}
\def\eea{\end{eqnarray}}

\def\prd{Phys. Rev. D}

\def\apss{Astrophysics and Space Science}

\def\physscr{Physica Scripta}

\def\sovast{Soviet Astronomy}

\begin{document}

\title{Gravitational capture cross-section in Zipoy-Voorhees spacetimes}

\author{Serzhan~Momynov}
\email[]{s.momynov@gmail.com}
\affiliation{Al-Farabi Kazakh National University, Al-Farabi Ave., 71, Almaty, 050040, Kazakhstan}
\affiliation{Satbayev University, Satbayev St., 22, Almaty, 050013, Kazakhstan}
\author{Kuantay~\surname{Boshkayev}}
\email[]{kuantay@mail.ru}
\affiliation{Al-Farabi Kazakh National University, Al-Farabi Ave., 71, Almaty, 050040, Kazakhstan}
\affiliation{National Nanotechnology Laboratory of Open Type, Almaty, 050040, Kazakhstana}
\affiliation{Institute of Nuclear Physics, Ibragimova, 1, Almaty, 050032, Kazakhstan}

\author{Hernando~\surname{Quevedo}}
\email[]{quevedo@nucleares.unam.mx}
\affiliation{Instituto de Ciencias Nucleares, Universidad Nacional Aut\'onoma de M\'exico, Mexico}
\affiliation{Dipartimento di Fisica and ICRA, Universit\`a di Roma “La Sapienza”, Roma, Italy}
\affiliation{Al-Farabi Kazakh National University, Al-Farabi Ave., 71, Almaty, 050040, Kazakhstan}

\author{Farida~\surname{Belissarova}}
\email[]{farida.belisarova@kaznu.kz}
\affiliation{Al-Farabi Kazakh National University, Al-Farabi Ave., 71, Almaty, 050040, Kazakhstan}

\author{Anar~\surname{Dalelkhankyzy}}
\email[]{dalelkhankyzy.d@gmail.com}
\affiliation{Kazakh National Women's Teacher Training University, Ayteke Bi, 99, Almaty, 050000, Kazakhstan}
\affiliation{Al-Farabi Kazakh National University, Al-Farabi Ave., 71, Almaty, 050040, Kazakhstan}

\author{Aliya~\surname{Taukenova}}
\email[]{aliya.tauken@gmail.com}
\affiliation{Al-Farabi Kazakh National University, Al-Farabi Ave., 71, Almaty, 050040, Kazakhstan}

\author{Ainur~\surname{Urazalina}}
\email[]{y.a.a.707@mail.ru}
\affiliation{Al-Farabi Kazakh National University, Al-Farabi Ave., 71, Almaty, 050040, Kazakhstan}
\affiliation{National Nanotechnology Laboratory of Open Type, Almaty, 050040, Kazakhstana}
\affiliation{Institute of Nuclear Physics, Ibragimova, 1, Almaty, 050032, Kazakhstan}

\author{Daniya~\surname{Utepova}}
\email[]{utepova\_daniya@mail.ru}
\affiliation{Abai Kazakh National Pedagogical University, Dostyk Ave., 13, Almaty, 050010, Kazakhstan}
\affiliation{Al-Farabi Kazakh National University, Al-Farabi Ave., 71, Almaty, 050040, Kazakhstan}

\date{\today}

\begin{abstract}
We consider geodesics of massive and massless  test particles in the gravitational field of a static and axisymmetric compact object described by the quadrupolar metric ($q$-metric), which is the simplest generalization of the Schwarzschild metric, containing an independent quadrupole parameter $q$. We analyze the effective potential profile and calculate the orbital parameters and capture cross-sections of test particles in this spacetime. 
Moreover, we derive the explicit expression for the escape angle of photons as a function of the quadrupole parameter. All the results reduce in the corresponding limit of vanishing quadrupole to the well-known case of the Schwarzschild spacetime. We argue that our results could be used to investigate realistic compact objects such as white dwarfs and neutron stars. 
\end{abstract}

\maketitle 

\section{Introduction}
Compact astrophysical objects, such as white dwarfs, neutron stars and black holes, are characterized by several phenomena, which emerge due to the dynamics of test particles moving in their vicinity. As a consequence, by analyzing the geodesics of the particles, one can infer information about the parameters and physical properties of the compact objects. To describe the geometry around compact objects, we can  use multipole moments that determine the structure of the spacetime uniquely \cite{beig1981multipole,van1985rotation,kundu1981analyticity,quevedo1990multipole}. In the case of vacuum spherically symmetric configurations, the Schwarzschild solution 
corresponds to the monopole moment. For compact objects, however, 
we should use higher multipole moments to account for the deviation from spherical symmetry. The simplest generalization of the Schwarzschild metric involving a quadrupole parameter is the Zipoy-Voorhees metric \cite{zipoy1966topology, voorhees1970static} (also known as $q$-metric, $\gamma$-metric, and $\delta$-metric).

Geodesics of the $q$-metric \cite{2000IJMPD...9..649H}, a stationary generalization of the static $q$-metric \cite{2014GrCo...20..252T}, and the motion of massive and massless test particles depending on the quadrupole parameter have been widely studied in the literature \cite{2016PhRvD..93b4024B}. 
The orbital parameters of test particles in accretion disks such as angular velocity $\Omega$, total energy $E$, angular momentum $L$, and radius of the innermost stable circular orbit $r_{ISCO}$ have been calculated as functions of  the mass and quadrupole parameters \cite{2016PhRvD..93b4024B}. In addition, the optical characteristics of accretion disks in the field of the static $q$-metric has been explored in \cite{2021PhRvD.104h4009B}.

The gravitational capture cross-section of test particles with non-zero and zero rest mass has been studied in the gravitational field of Schwarzschild black holes \cite{1975ctf..book.....L, 1985bhwd.book.....S,1988SvA....32..456Z},  Schwarzschild--Tangherlini black holes \cite{2021Univ....7..307A}, and the Myers-Perry metric \cite{2008PhRvD..77j4026G}. The escape of  massless and massive particles from the Kerr--Sen black hole to spatial infinity \cite{2021NuPhB.96415313Z} and  the impact of the leading coefficient of the parameterized line element of the spherically symmetric, static black hole on the capture cross-section of massless and massive particles have also been studied in \cite{2021Galax...9...65T}.

Capture cross-sections were obtained  for  photons (or uncharged ultra-relativistic particles) and for uncharged particles, which are slowly moving  at infinity in the field of a spherically symmetric, charged black holes \cite{1994CQGra..11.1027Z} and  compact charged body \cite{1992TMP....90...97Z}. 

Analytic expressions for the capture cross-sections of magnetized particles in the field of a Schwarzschild black hole and of a braneworld black hole immersed in an external magnetic field  were derived in \cite{2014PhyS...89h4008A} and \cite{2011Ap&SS.335..499R}, respectively. 

The main purpose of this work is to understand the influence of the quadrupole moment on  the capture cross-section for massless  and massive test particles in the gravitational field of the $q$-metric. 

This paper is organized as follows.
In Sec. \ref{sec:massive}, we review the main properties of the $q-$metric, analyze the corresponding geodesic equations to obtain the effective potential for massive test particles, and calculate and analyze the behavior of the capture cross-section. We also use our results to investigate the efficiency of the source of gravity to convert mass into radiation. Section \ref{sec:massless} contains the results obtained for the capture cross-section in the case of massless particles. In Sec. \ref{sec:escape}, we derive the explicit expression for the capture angle and show the consistency of our results in the limiting case of the Schwarzschild spacetime. Finally, in Sec. \ref{sec:con}, we summarize the main results of our work. 

\section{Motion of massive test particles}
\label{sec:massive}

The simplest generalization of the Schwarzschild metric with a quadrupole parameter is the $q$-metric. In spherical coordinates, the $q$-metric has the following form: 
\begin{equation}
\begin{gathered}
    ds^{2}=-f^{1+q}dt^{2}+f^{-q}\Bigg[g^{-q(2+q)}\left( \frac{dr^{2}}{f}+r^{2}d\theta^{2}\right)\\+r^{2}\sin^{2}\theta d\phi^{2}\Bigg],
\end{gathered}
\end{equation}
where $M$ and $q$ are the mass and quadrupole parameters of the source, respectively, and
\be 
f=f(r)=1-\frac{2M}{r}, \ee
\be g=g(r,\theta)=1+\frac{M^{2}\sin^{2}\theta}{r^{2}f}.
\ee 
In the limiting case $q=0$, we recover the spherically symmetric Schwarzschild geometry.

To investigate the motion of test particles in the gravitational field described by the $q-$metric, it is convenient to consider the Lagrange function 
\begin{equation}\label{eq1}
\begin{gathered}
    2\mathcal{L}= -f^{1+q}\dot{t}^{2}+f^{-q} \Bigg[g^{-q(2+q)}\left( \frac{\dot{r}^{2}}{f}+r^{2}\dot{\theta}^{2}\right)\\+r^{2}\sin^{2}\theta \dot{\phi}^{2}\Bigg] \, ,
\end{gathered}
\end{equation}
where 
$\dot{t}\equiv dt/ d\lambda=p^{t}$, $\dot{r}\equiv dr/ d\lambda=p^{r}$, $\dot{\theta}\equiv d\theta/ d\lambda=p^{\theta}$ and $\dot{\phi}\equiv d\phi/ d\lambda=p^{\phi}$.
We choose the affine  parameter as $\lambda=\tau/m$ for a particle with mass $m$.

The Euler-Lagrange equations 
\begin{equation}\label{eq2}
    \frac{d}{d \lambda}\left(\frac{\partial \mathcal{L}}{\partial \dot{x}^{\alpha}}\right)-\frac{\partial \mathcal {L}}{\partial x^{\alpha}}=0, ~~~~ x^{\alpha}=(t,r,\theta,\phi),
\end{equation}
for the coordinates $t, \theta, \phi$ are given by:
\begin{equation}\label{eq3}
    \frac{d}{d\lambda}\left(f^{1+q}\dot{t}\right)=0 ,
\end{equation}
\begin{equation}\label{eq4}
\begin{gathered}
    \frac{d }{d\lambda} \left(r^{2} f^{-q}g^{-q(2+q)}\dot{\theta}\right) = -f^{-q} \cos\theta\sin\theta\biggl[ \dot{\phi}^{2}r^{2}\\-M^{2}q(2+q)\left( \frac{\dot{r}^{2}}{f}+\dot{\theta}^{2}r^{2}\right)\frac{g^{-1-q(2+q)}}{r^{2}f} \biggr],
\end{gathered}
\end{equation}
\begin{equation}\label{eq5}
    \frac{d}{d\lambda}\left(r^{2}\sin^{2}\theta f^{-q}\dot{\phi}\right)=0
\end{equation}

To find the equation for the coordinate $r$, it is convenient to use the normalization condition for the four-dimensional momentum, 
\begin{equation}\label{eq6}
     g_{\alpha\beta}p^{\alpha}p^{\beta}=-m^{2} , 
 \end{equation}
so that in this case $\mathcal{L}$ is equal to $-m^{2}/2$.

From Eq.\eqref{eq4}, we can conclude that  a particle moving on the equatorial plane ($\theta=\pi/2$) will remain on the same plane. This is a consequence from the fact that the $q-$metric is invariant under  reflections with respect to the equatorial plane. 

For $\theta=\pi/2$, from Eqs. \eqref{eq4} and  \eqref{eq5} it follows that 
\begin{equation}\label{eq8}
    -p_{t}\equiv f^{1+q}\dot{t}=E
\end{equation} 
\begin{equation}\label{eq7}
     p_{\phi}\equiv r^{2}f^{-q}\dot{\phi}=L ,
\end{equation}
where $E$ and $L$ are real constants. 

For a particle with non-zero mass, new parameters for the energy $\tilde E$ and angular momentum  $\tilde L $ can be defined as 
\begin{equation}\label{eq9}
    \Tilde{E}=\frac{E}{m},~~~~~\Tilde{L}=\frac{L}{m}.
\end{equation}
If we take into account that $\lambda=\tau/m$, from Eqs. \eqref{eq4}--\eqref{eq6}, we obtain 
\begin{equation}\label{eq10}
\begin{gathered}    
    \left(\frac{dr}{d\tau}\right)^{2}=g^{q(2+q)}\left[\Tilde{E}^{2}-
    U^2\right]
\end{gathered}
\end{equation}
where $U$ is the effective potential defined by
\begin{equation}
    U^2=f^{1+q}\left(1+\frac{\Tilde{L}^{2}}{r^{2}f^{-q}}\right) 
\end{equation}
Moreover, the remaining components of the four-velocity can be represented as
\begin{equation}\label{eq12}
    \frac{dt}{d\tau}=\frac{\Tilde{E}}{f^{1+q}} ,
\end{equation}
\begin{equation}\label{eq11}
    \frac{d\phi}{d\tau}=\frac{\Tilde{L}}{r^{2}f^{-q}} .
\end{equation}

\begin{figure}
\centering
\includegraphics[width=\linewidth]{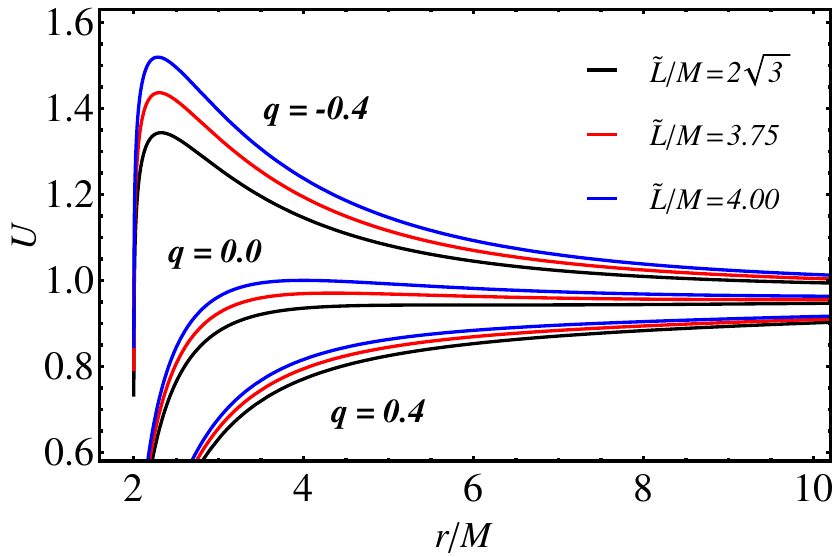}
\caption{Effective potential profile for particles with non-zero rest mass and different values of angular momentum $\tilde L$, moving in orbit relative to the gravitational field of a static naked singularity generated
by a mass $M$ and quadrupole moment $q$=-0.4, 0, 0.4}
\label{fig:mpr1}
\end{figure}

\begin{figure}
\centering
\includegraphics[width=\linewidth]{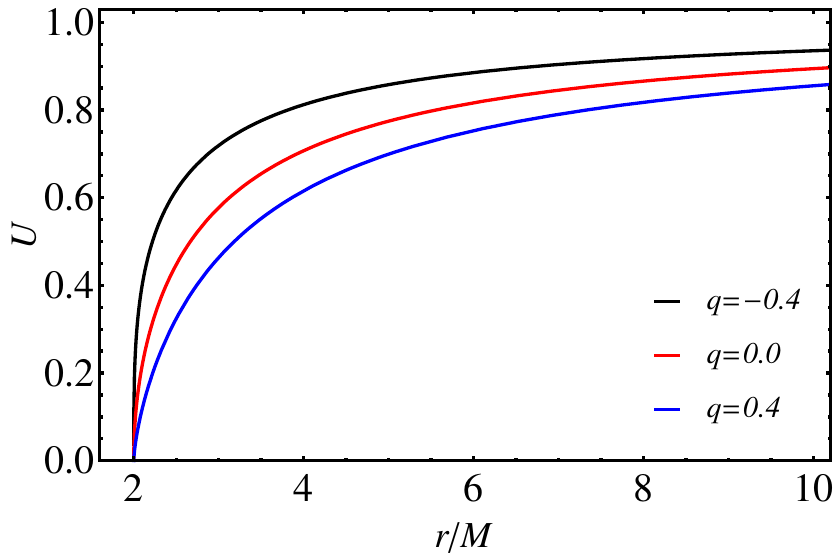}
\caption{Effective potential profile for particles with non-zero rest mass and  angular momentum  $\tilde L =0$, moving in orbit relative to the gravitational field of a static naked singularity generated
by a mass $M$ and quadrupole moment $q$=-0.4, 0, 0.4}
\label{fig:mpr2}
\end{figure}

Figs. \ref{fig:mpr1} and \ref{fig:mpr2} show 
the dependence of the effective potential $U$  from the normalized radial coordinate $r/M$ for various values of the orbital angular momentum $\Tilde{L}$. The points of local minima correspond to the radius of stable circular orbits.
The motion of test particles on the equatorial plane of the $q-$metric is essentially determined by Eqs.(\ref{eq10}), (\ref{eq12}), and (\ref{eq11}).

\subsection{
Cross-sections}
\label{sec:cross}

The capture cross-section of particles is defined as 
\begin{equation}\label{eq13}
    \sigma_{capt}=\pi b_{max}^{2},
\end{equation}
where $b_{max}$ is the maximum impact parameter of the captured particle.  
From Eqs.~\eqref{eq10} and   Eq.~\eqref{eq11}, we obtain 
\begin{equation}\label{eq14}
\begin{gathered}
        \frac{1}{r^{4}}\left(\frac{dr}{d\phi}\right)^{2}
        =\frac{1}{\Tilde{L}^{2}f^{2q}}\left(1+\frac{M^{2}}{r^{2}f}\right)^{q(2+q)} \left[\Tilde{E}^{2}-U^{2} \right].
\end{gathered}
\end{equation}
From the condition
\be 
    \frac{ d U^{2}}{ d r}=0,
\ee
we can determine the value of the angular momentum as
\be
\Tilde{L}^{2}=f^{-q}\frac{M(1+q)r^{2}}{r-M(3+2q)}
\label{eq15}
\ee
Furthermore, the circular orbit condition  $\dot{r}=0$ 
is equivalent to 
\begin{equation}\label{eq16}
     U^{2}=\Tilde{E}^{2}, ~~~~~\Tilde{E}^{2}=f^{1+q}\frac{r-M(2+q)}{r-M(3+2q)}.
\end{equation}

It is interesting to note that the conditions $d^2U/dr^2=0$ and $d\tilde{L}/dr=0$, or equivalently $d\tilde{E}/dr=0$, lead to the radius of the innermost stable circular orbit $r_{ISCO}$, which for the $q$-metric is
\begin{equation}
    \frac{r_{ISCO}}{M}=4+3q+\sqrt{4+10q+5q^2}.
\end{equation}
Requiring that the square root is positive we obtain two conditions $q<-(1+1/\sqrt{5})\approx -1.447$ and  $q>-(1-1/\sqrt{5})\approx -0.553$.  Hence, to get real values for  $r_{ISCO}$, we choose the condition $q>-0.553$.

Another interesting quantity is the efficiency of the source to convert mass into radiation $\eta$, which is defined as
\begin{equation}
    \eta=\left[1-\tilde{E}(r_{ISCO})\right]\times 100\%
\end{equation}

\begin{figure}
\centering
\includegraphics[width=\linewidth]{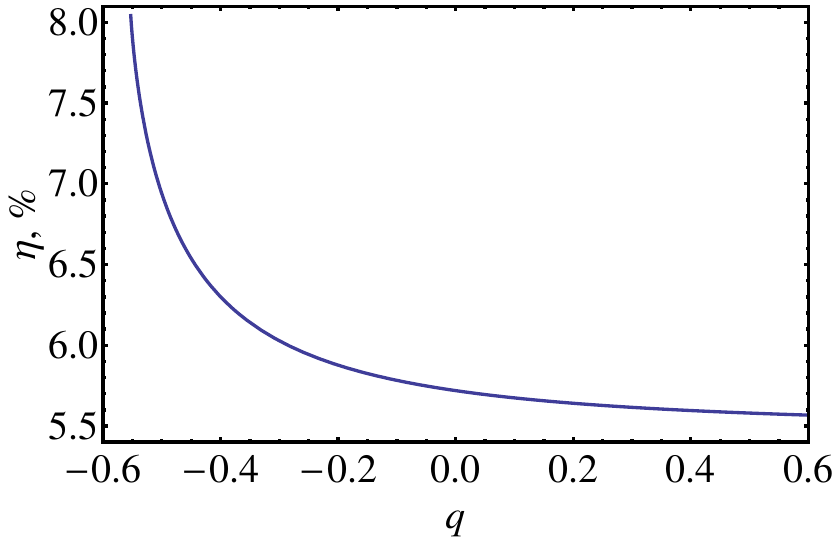}
\caption{Efficiency of converting matter into radiation in the $q$-metric.}
\label{fig:eta}
\end{figure}

In Fig.~\ref{fig:eta}, we show the efficiency as a function of $q$ when $M=1$. As one can see, the maximum efficiency 8.0417$\%$ is achieved for $q=1/\sqrt{5}-1\approx-0.553$. For $q=0$, we obtain the classical result for the Schwarzschild black hole $\eta=5.719\%$. As $q\rightarrow\infty$, $\eta\rightarrow 5.478\%$. We can conclude that the efficiency in the $q-$spacetime is larger (smaller) than the Schwarzschild one for prolate (oblate) sources.

\begin{figure*}
{\hfill
\includegraphics[width=0.46\hsize,height=0.28\hsize,clip]{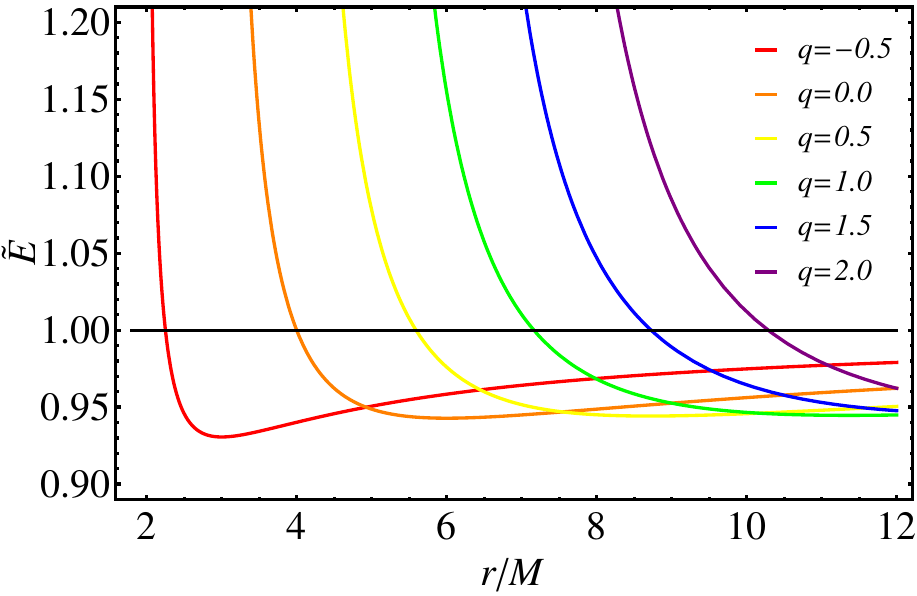}\hfill
\includegraphics[width=0.44\hsize,clip]{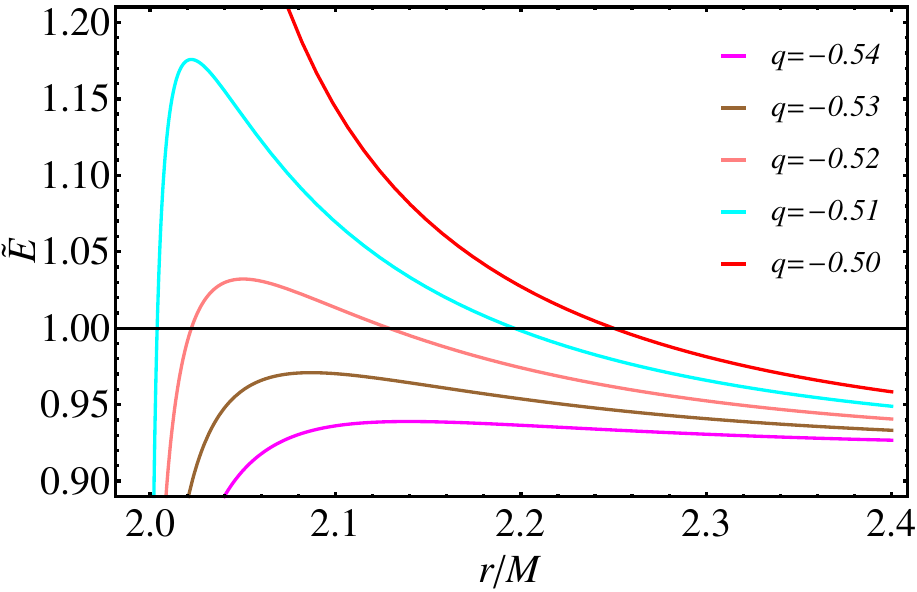}\hfill}
\caption{Energy of test particles versus dimensionless radial coordinate for different values of the quadrupole parameter.}
\label{fig:energy}
\end{figure*}

For parabolic trajectories, the condition $\Tilde{E}=1$ yields the radius of minimal approach to the central object. Since the expression for $\Tilde{E}$ is in the form of a transcendent equation, it is not possible to derive the values of $r$ analytically.  However, for small values of $q$ it is always possible to find an approximate analytical solution. Therefore, expanding  $\Tilde{E}^{2}$ in a Taylor series and  retaining only linear terms in $q$, we obtain
\begin{equation}\label{eq17}
\begin{gathered}
   \Tilde{E}^{2}\approx f\frac{r-2M}{r-3M}\\+qf\biggl[\frac{(r-M)M}{(r-3M)^{2}}+\frac{r-2M}{r-3M}\ln f\biggr]+O(q^2).
\end{gathered}
\end{equation}
Furthermore, the solution for $r$ can be found in the form $r=4M+q z(M)$, where $z(M)$ is the sought function. By plugging $r$ in the above equation, then expanding it in a  Taylor series, and equating it again to 1, we obtain the following expression
\begin{equation}\label{eq18}
   \left( \frac{3}{2}-\frac{z(M)}{4M}-\ln2 \right)q=0,
\end{equation}
which gives
\begin{equation}\label{eq19}
    z(M)=2M(3-2\ln2).
\end{equation}

Finally, the  approximate radius of 
 marginally bound orbits can be expressed  in the form 
 \begin{equation}
  r_{mb}=4M+2Mq(3-2\ln2).   
 \end{equation}
Similarly, a Taylor expansion of  $\Tilde{L}$ in powers of $r$ evaluated at $r=r_{mb}$  yields 
\begin{equation}\label{eq20}
    \Tilde{L}^{2}\approx 16M^{2}(1+3q\ln2)
\end{equation}

In Fig.~\ref{fig:energy}, we present the orbital energy of test particles as a function of the radial distance from the center of gravity. In the left panel, for relatively large values of $q$, we see that the energy curve always crosses the horizontal line $\tilde{E}=1$. However, in the right panel, we see that for small values of $q$ there are cases in which the condition $\tilde{E}=1$ is violated, implying  that there is a minimum value of $q$ for which one can have parabolic geodesics in the $q$-metric. A numerical evaluation indicates that the condition $\tilde{E}=1$ is fulfilled if the  quadrupole parameter satisfies that $q\geq-0.524384$.

\begin{figure}
\centering
\includegraphics[width=\linewidth]{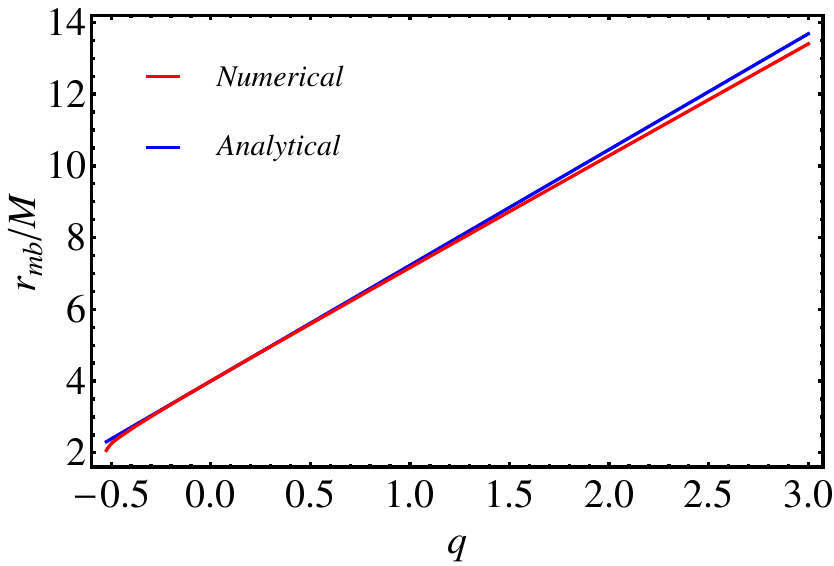}
\caption{Radii of marginally bound orbits versus the quadrupole parameter.}
\label{fig:rmb}
\end{figure}

The radii of marginally bound orbits as a function of $q$ in the $q$-metric is illustrated in Fig.~\ref{fig:rmb}. As one may notice the approximate analytical solution and the exact numerical solution of the condition   $\tilde{E}=1$ (for $M=1$) show similar behavior near zero. For larger and smaller values of $q$ they diverge, as expected.

The impact parameter is defined by $b=\Tilde{L}/v_{\infty }$, and the capture cross-section by
$\sigma=\pi b^{2}=\pi\Tilde{L}^{2}/v^{2}_{\infty }$.
Thus, for the $q$-metric the capture cross-section is given by:
\begin{equation}\label{eq21}
    \sigma=\pi\frac{16M^{2}}{v^{2}_{\infty}}(1+3q\ln2)
\end{equation}
when $q$ goes to zero, the formula represents the result for the Schwarzschild metric.

\begin{figure}
\centering
\includegraphics[width=\linewidth]{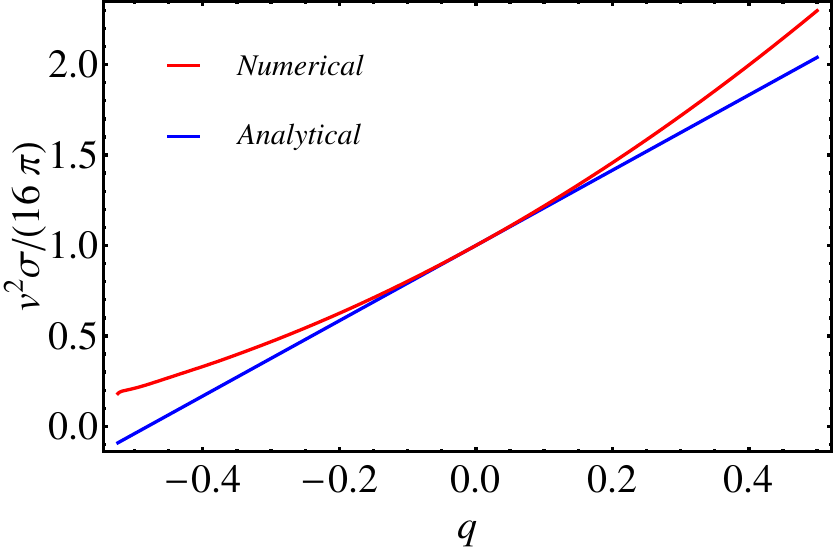}
\caption{Normalized gravitational capture cross-section versus quadrupole parameter.}
\label{fig:sigma}
\end{figure}

The normalized capture cross-section as a function of quadrupole parameter is depicted in Fig.~\ref{fig:sigma}. The approximate analytical and the exact numerical solution produce similar outcomes close to $q=0$. However, at large and small $q$ the results diverge as expected. Moreover, as mentioned above, the quadrupole parameter must be $q\geq-0.524384$, which results in a minimum positive capture cross-section from the exact numerical analyses, whereas the approximate analytical solution yields non-physical results.

\section{Motion of massless test particles}
\label{sec:massless}

For a photon ($m$=0) equations \eqref{eq4}--\eqref{eq6} take the form
\begin{equation}\label{eq22}
    \frac{dt}{d\lambda}=\frac{E}{f^{1+q}}
\end{equation}
\begin{equation}\label{eq23}
    \frac{d\phi}{d\lambda}=\frac{L}{r^{2}f^{-q}}
\end{equation}
\begin{equation}\label{eq24}
    \left(\frac{dr}{d\lambda}\right)^{2}
    =g^{q(2+q)}\left[E^{2}-U_{ph}^2\right], 
\end{equation}
respectively, where $U_{ph}$ is the effective potential for massless particles defined by
\begin{equation}
 U_{ph}^2=\frac{\tilde{L}^{2}}{r^{2}}f^{1+2q} .  
 \label{eq:effpotph}
\end{equation}
For simplicity, we  write the equation  for the case $m$=0 by applying the transformation $\lambda_{new}=L\lambda$ and $b=L/E$. Dropping the subscript ``new'', we have
\begin{equation}\label{eq25}
    \frac{dt}{d\lambda}=\frac{1}{bf^{1+q}}
\end{equation}
\begin{equation}\label{eq26}
    \frac{d\phi}{d\lambda}=\frac{1}{r^{2}f^{-q}}
\end{equation}
\begin{equation}\label{eq27}
    \left( \frac{dr}{d\lambda}\right)^{2}=g^{q(2+q)}\left( \frac{1}{b^{2}}-\frac{f^{1+2q}}{r^{2}}\right)
\end{equation}
Photon orbits  can be obtained from the effective potential Eq.~\eqref{eq:effpotph}.  The critical point for $r$ is given by
\begin{equation}\label{eq:rph}
\frac{r_{Ph}}{M}=(3+2q)
\end{equation}
which corresponds to the photosphere radius. Then, using the condition $\dot{r}=0$, $b_{c}$ takes form
\begin{equation}\label{eq29}
    b_{c}^{2}=\frac{\left(3+2q\right)^{3+2q}}{\left(1+2q\right)^{1+2q}}M^2
\end{equation}
and the cross-section becomes
\begin{equation}\label{eq30}
    \sigma=\pi b_{c}^{2}=\frac{\left(3+2q\right)^{3+2q}}{\left(1+2q\right)^{1+2q}}\pi M^2 .
\end{equation}
The limiting case $q$=0 reproduces the known result for the Schwarzschild metric.

\begin{figure}
\centering
\includegraphics[width=\linewidth]{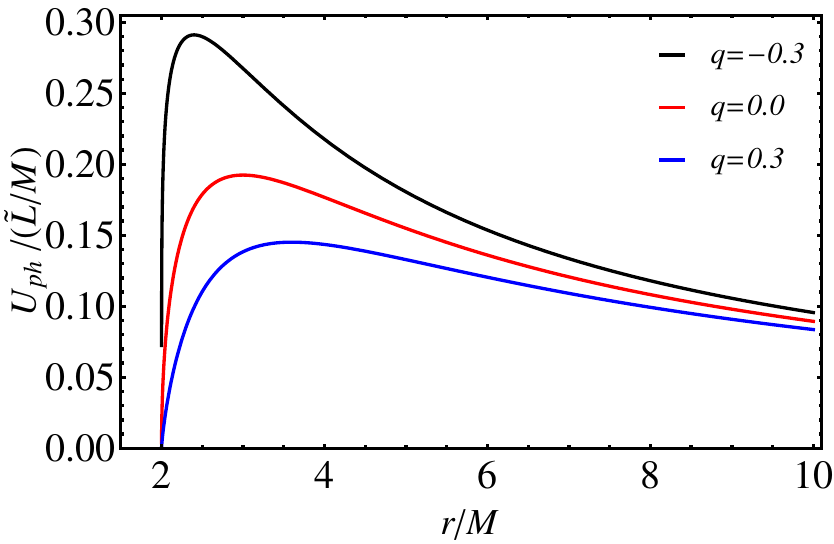}
\caption{Effective potential profile for a particle with zero rest mass moving along an orbit in the gravitational field of the $q$-metric. Here, the quadrupole parameter has the values $q$=-0.3, 0, 0.3.}
\label{fig:mpr3}
\end{figure}
One can see from here that the cross-section can be physical only when $q>-0.5$.
{ For the Schwarzschild metric it is known that  if the impact parameter $b<3\sqrt{3}M$, then the particle  with zero rest mass  is captured by the black hole. }

Referring to Fig.~\ref{fig:mpr3}, we can say that as the value of $q$ increases, the value of the impact parameter decreases.
The particle with zero rest
mass will be captured by the black hole with a lower value of the impact parameter.

\begin{figure}
\centering
\includegraphics[width=\linewidth]{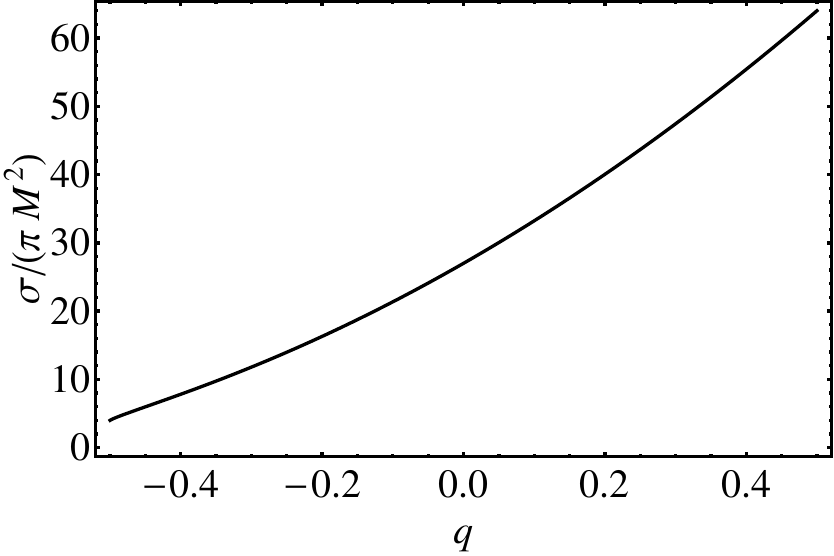}
\caption{Capture cross-section versus quadrupole parameter for photons in the $q$-metric.}
\label{fig:sigmaqphoton}
\end{figure}
In Fig.~\ref{fig:sigmaqphoton} we show the capture cross-section normalized by $\pi M^2$ for massless particles. As $q$ increases, the cross-section increases as well.

\section{Escape versus capture as a function of the propagation direction}
\label{sec:escape}

To calculate the observed radiation of particles near a black hole, it is necessary to know the direction of motion measured by a static nearby observer, for which a photon emitted at radius $r$ can escape to infinity. It can be seen that at 
$ 
r\geq(3+2q)M
$ 
the photon escapes unless $v^{\hat{r}}>0$ otherwise $v^{\hat{r}}<0$ and $b>\left(3+2q\right)^{3/2+q}M/\left(1+2q\right)^{1/2+q}$. Let us introduce the escape angle $\psi$ between the direction of motion and the radius ($\boldsymbol{|{v}=1|}$) (for more details see \cite{1985bhwd.book.....S,misner1973gravitation}):
\begin{equation}\label{eq31}
    v^{\hat{\phi}}=\sin \psi, ~~~~~~v^{\hat{r}}=\cos \psi  
\end{equation}
Since the equation of motion \eqref{eq1} corresponds to a cyclic coordinate, locally the orthonormal system is represented by the locally measured energy: 
\begin{equation}\label{eq32}
\begin{gathered}
    E_{local}\equiv p^{\hat{t}}=-p_{\hat{t}}=-\boldsymbol{\Vec{p}} \cdot \boldsymbol{\Vec{e}_{\hat{t}}}=\\=-\boldsymbol{\Vec{p}}f^{-\frac{1+q}{2}}\boldsymbol{\vec{e}_{\hat{t}}}=-f^{-\frac{1+q}{2}}p_{t}
\end{gathered}
\end{equation}
\begin{equation}\label{eq33}
    E=f^{\frac{1+q}{2}}E_{local}
\end{equation}
When $r\rightarrow\infty$ ; $E_{local}\rightarrow E$. Then, $E$ is related to the $E_{local}$ with a multiplicative term that determines the redshift.
Taking into account Eqs.\eqref{eq33},\eqref{eq7} and $b_{c}=L/E$, the tangential velocity of the particle is determined by
\begin{equation}\label{eq34}
\begin{gathered}
        v^{\hat{\phi}}=\frac{p_{\phi}}{rE_{local}}=\frac{p_{\phi}f^{\frac{1+q}{2}}}{rE}=\frac{l}{rE}f^{\frac{1+q}{2}}=\frac{b_{c}}{r}f^{\frac{1+q}{2}}.
\end{gathered}
\end{equation}
A photon moving in the direction of the black hole will escape from it if
\begin{equation}\label{eq35}
\begin{gathered}
\sin\psi > \frac{M(3+2q)^{3/2+q}}{r\left(1+2q\right)^{1/2+q}}f^{(1+q)/2}.
\end{gathered}
\end{equation}
\begin{figure}
\centering
\includegraphics[width=\linewidth]{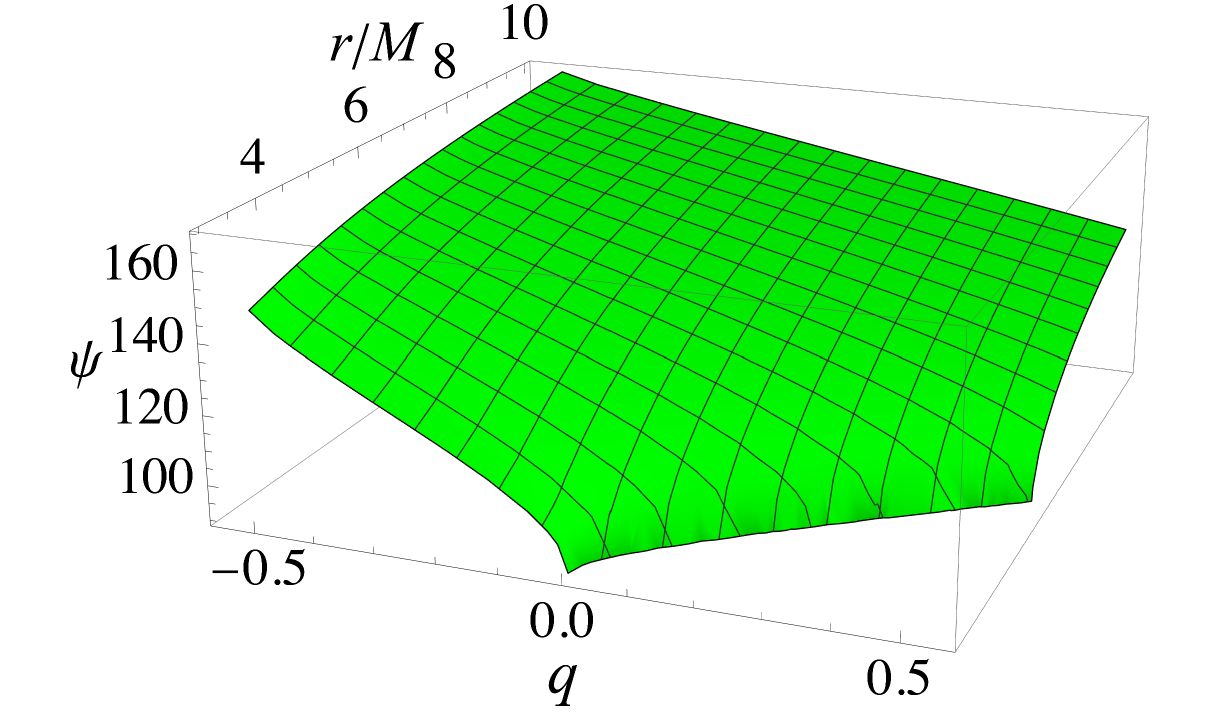}
\caption{3D plot of $\psi$ as a function of $q$ and $r/M$.}
\label{fig:psiplot3D}
\end{figure}
In Fig.~\ref{fig:psiplot3D},
we show the behavior of the angle $\psi$ in terms of the radial coordinate and the quadrupole parameter. This shows clearly that the value of $\psi$ decreases as $q$ and $r/M$ increase.

For a Schwarzschild spacetime  at $r=6M$, 
we obtain 
\begin{equation}
    \sin\psi > \frac{1}{\sqrt{2}}
\end{equation}
so that the escape angle should be $\psi<135^{0}$. In turn,  for Zipoy-Voorhees spacetimes at $r=6M$, we obtain 
\begin{equation}\label{eq36}
\begin{gathered}
\sin\psi > \frac{(3+2q)^{3/2+q}}{6\left(1+2q\right)^{1/2+q}}\left(\frac{2}{3}\right)^{(1+q)/2}.
\end{gathered}
\end{equation}
In Table~\ref{tab:sinpsi}, we present the results obtained for the escape angle for different specific values of the radial distance in spacetimes with and without quadrupole.
%
%

In the limiting case of small values of $q$, the   expression for the escape angle reduces to 

\begin{equation}\label{38}
\begin{gathered}
\psi < 135^0 - \frac{90^0\ln 6}{\pi}q+O(q^2)\\\approx 135^0 -51.33^0q +O(q^2), 
\end{gathered}    
\end{equation}
which shows clearly how the quadrupole parameter changes the value of $\psi$. 

The explicit expression of the scape angle for $r=r_{ISCO}$  is given in Table \ref{tab:sinpsi}, which in the limiting case of small $q$ becomes

\begin{equation}\label{37}
\begin{gathered}
\psi < 135^0 + \frac{45^0(11-\ln{1679616})}{4\pi}q+O(q^2)\\\approx 135^0 -11.939^0q+O(q^2),    
\end{gathered}    
\end{equation}
%
%
%
%
%
%

\begin{table}
\centering
\begin{tabular}{ccc}
\hline
\hline
    &           &                                      \\
$r$    & $\sin\psi>$ & $\sin\psi>$\\
    &  for $q\neq0$  &       for $q=0$                                       \\
\hline
   &          &                                        \\
$3M$ & $\frac{(3+2q)^{3/2+q}}{3^{(3+q)/2}(1+2q)^{1/2+q}}$  &  1                                      \\
    &       &                                           \\
$r_{Ph}$ &  $\left(\frac{3+2q}{1+2q}\right)^{q/2}$   & 1   \\
    &                                                  \\
$6M$ & $\frac{(3+2q)^{3/2+q}}{6(1+2q)^{1/2+q}}\left(\frac{2}{3}\right)^{(1+q)/2}$ & $1/\sqrt{2}$\\
    &                                                 \\
$r_{ISCO}$ & $\frac{(3+2q)^{3/2+q}}{(1+2q)^{1/2+q}}\frac{(2+3q+\sqrt{4+10q+5q^2})^{(1+q)/2}}{(4+3q+\sqrt{4+10q+5q^2})^{(3+q)/2}}$ &  $1/\sqrt{2}$  \\
    &          &                                        \\
\hline
\end{tabular}
\caption{Expressions of $\sin\psi$ at different radii.}
\label{tab:sinpsi}
\end{table}

\begin{figure*}
{\hfill
\includegraphics[width=8cm]{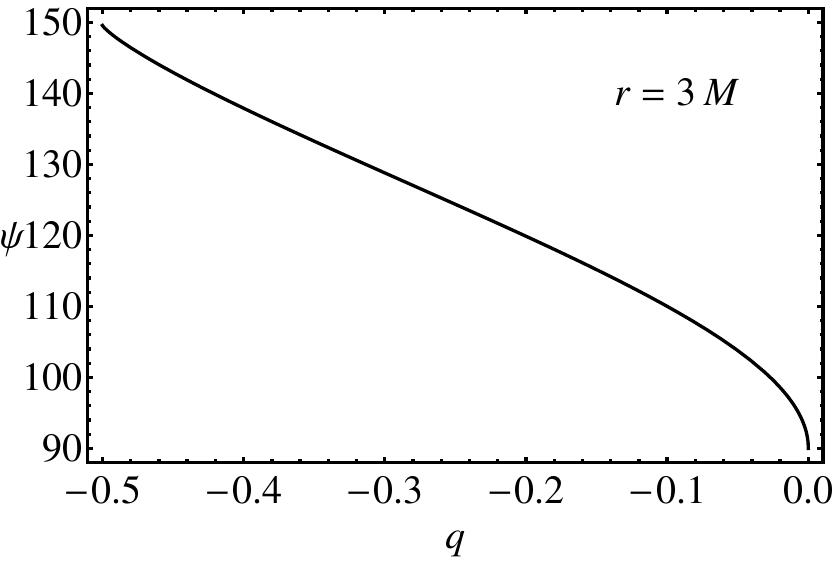}\hfill
\includegraphics[width=8cm]{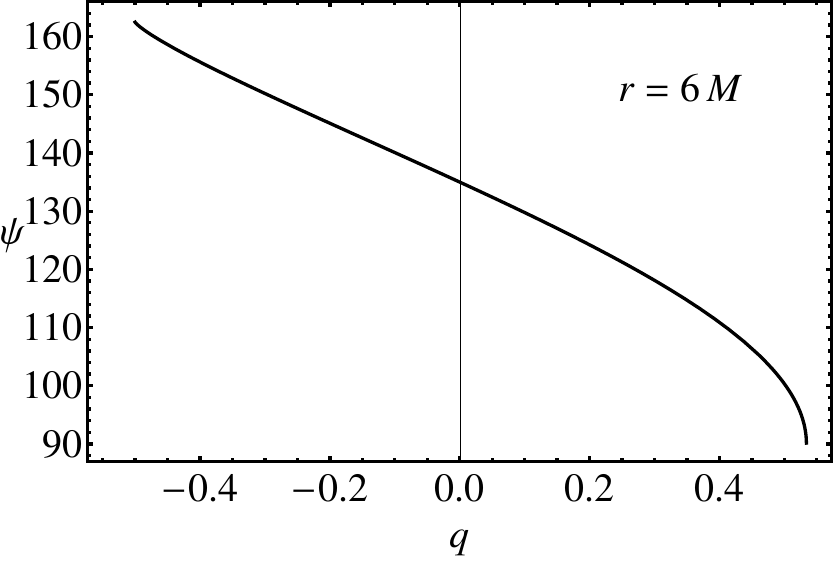}\hfill} 

{\hfill
\includegraphics[width=8cm]{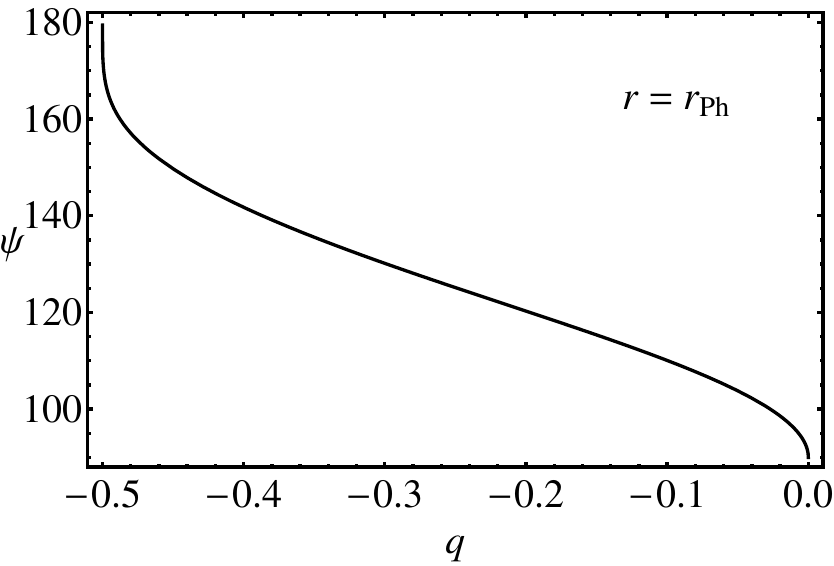}\hfill
\includegraphics[width=8cm]{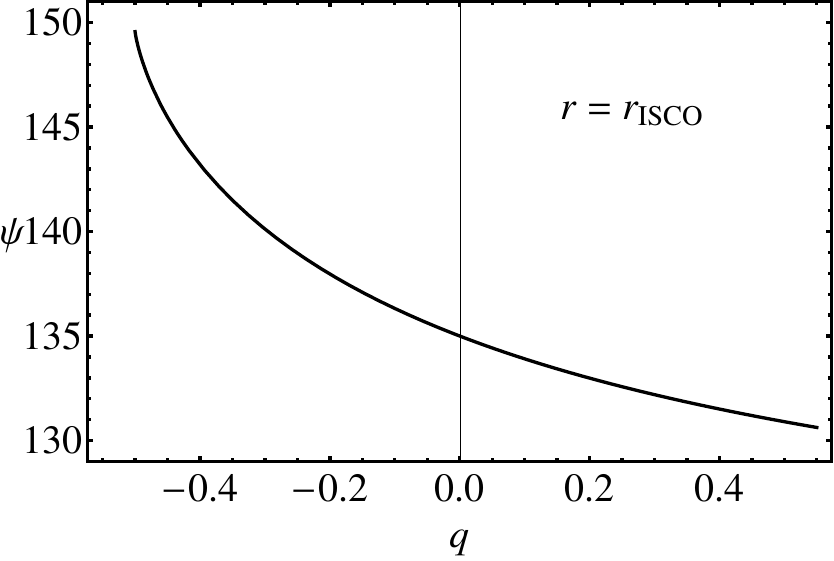}\hfill} 

\caption{
$\psi$ versus $q$ at different radii.
Left panel. 
Top: $r=3M$. 
Bottom: $r=r_{Ph}$.  
Right panel. 
Top: $r=6M$. 
Bottom: $r=r_{ISCO}$. 
}\label{fig:psiq} 
\end{figure*}

{In Fig.~\ref{fig:psiq}, we plot $\psi$ versus $q$ for different fixed values of the radial distance $r$. As one may notice, for a  fixed of $r$, $\psi$ decreases with increasing $q$.}

\section{Conclusions}
\label{sec:con}

In this work, we investigated the  gravitational capture cross-sections of massive and massless test particles on the equatorial plane of  the Zipoy-Voorhees spacetime. All the obtained results in the limiting case $q \rightarrow 0$ reduce to the expressions already known for the Schwarzschild metric, which is an indication of the correctness of our results. 

Our approach consists in using the geodesic equations for deriving the  effective potential for test particles with  zero and non-zero rest masses, moving in the gravitational field represented by the $q-$metric. We analyzed the behaviour of the effective potential for different values of the quadrupole parameter $q$ and obtained consistent results that reduce to the Schwarzschild case in the corresponding limit.  

 The obtained results could be useful in describing capture cross-sections of realistic compact objects with quadrupole moment, namely white dwarfs and neutron stars.  
 It would be interesting to extend this work for geodesics in the field of rotating and deformed compact objects. This issue will be addressed in future studies.

\section*{Acknowledgements}
KB is supported by Grant No. AP19680128; AU is supported by Grant No. BR21881941, DU is supported by Grant No. AP22682939, all from the Science Committee of the Ministry of Science and Higher Education of the Republic of Kazakhstan.
The work of HQ was supported by UNAM PASPA-DGAPA.

\end{document}